\documentclass[twocolumn]{article}
\usepackage[top=1in, bottom=1in]{geometry}
\usepackage{txfonts}

\usepackage[dvips]{graphicx}
\usepackage[normalem]{ulem}

\usepackage{hyperref}
\usepackage{url}
\usepackage{cite}

\usepackage{authblk}

\usepackage[usenames, dvipsnames]{color}

\usepackage{capt-of}
\usepackage{etoolbox}
\makeatletter
\apptocmd\@maketitle{{\myteaser{}\par}}{}{}
\makeatother

\newcommand\myteaser{%
\centering
    \includegraphics{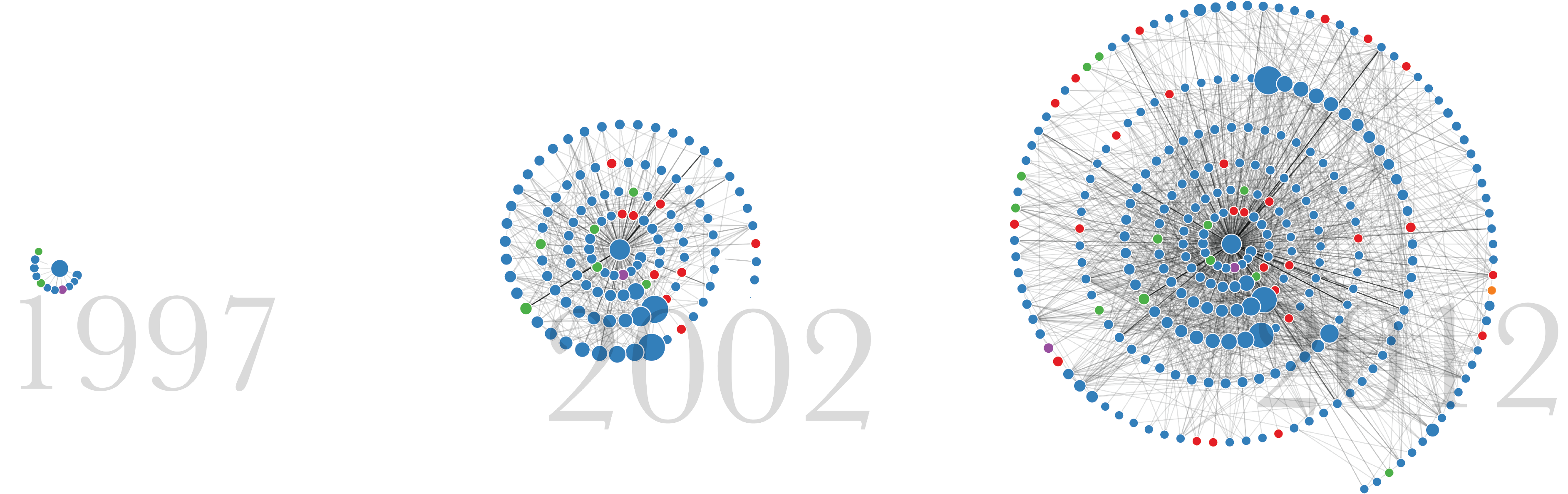}
\captionof{figure}{Three screenshots of the network animation as it builds over time.}
\label{fig:teaser}
}
\title{Leveraging Citation Networks to Visualize Scholarly Influence Over Time}

\date{October, 2016}

\author[1]{Jason Portenoy}
\author[1]{Jessica Hullman}
\author[1]{Jevin D. West}
\affil[1]{The Information School, University of Washington}

\setkeys{Gin}{width=\linewidth,keepaspectratio}

\begin{document}



\twocolumn[
\begin{@twocolumnfalse}
\maketitle

\begin{abstract}

Assessing the influence of a scholar's work is an important task for funding organizations, academic departments, and researchers. Common methods, such as measures of citation counts, can ignore much of the nuance and multidimensionality of scholarly influence. We present an approach for generating dynamic visualizations of scholars' careers. This approach uses an animated node-link diagram showing the citation network accumulated around the researcher over the course of the career in concert with key indicators, highlighting influence both within and across fields. We developed our design in collaboration with one funding organization---the Pew Biomedical Scholars program---but the methods are generalizable to visualizations of scholarly influence. We applied the design method to the Microsoft Academic Graph, which includes more than 120 million publications. We validate our abstractions throughout the process through collaboration with the Pew Biomedical Scholars program officers and summative evaluations with their scholars.


\end{abstract}
\vspace{12pt}
\end{@twocolumnfalse}
]

\section{Introduction}\label{introduction}

The scholarly literature forms a vast network that is connected through citations and footnotes.  This well-preserved system---through its billions of links---connects papers, authors, ideas and disciplines over centuries. The structure of this system can reveal where ideas have come from and where they might be going. Though De Solla Price first recognized the potential of citation networks for improving search, evaluation and discovery more than 50 years ago~\cite{de_solla_price_networks_1965}, realizing the potential of citation networks for conveying patterns in scholarship has been challenging.  
Recent advances in data access and scaling pave the way for increased focus on how to communicate the insight captured in citation networks.

One common scenario that calls for ways to accurately and efficiently convey citation network data is measuring scholarly influence. Funding agencies, hiring and promotion committees, and university leaders want to measure the impact of their scholars, but few tools sufficiently address this task.
There have been many proposed metrics for measuring influence (\cite{hirsch_index_2005}, \cite{yan_scholarly_2012}), but none suffice in capturing the full complexity of scholarly influence. For these aspects, it can be more effective to visualize the movement of ideas between papers via direct citations. There have been many attempts at mapping the scholarly literature using citation networks \cite{cobo_science_2011}, however, most of these attempts view science at the aggregate, disciplinary level. For this paper, we focus at the local view---at the level of an individual author---with an interest in depicting the \emph{influence} of this author over time.  
Specifically, we are interested in temporal, author-level citation networks in which the nodes represent papers that cite the work of a particular scholar. 

A number of different parties have an interest in looking at the
influence of scholarly work and individual scholars. Funding organizations---including nonprofits
and government agencies such as the National Institute of
Health---collectively spend billions of dollars annually to fund
research. These organizations are continually faced with the question of
how best to evaluate the impact that the funding has had. University
departments tasked with hiring and promotion decisions must evaluate the
impact of research as well. Many scholars are interested in looking at
their own influence as a means of self reflection, or at other scholars in their field.

The primary contribution of this paper is a broadly accessible, automated, data-driven approach to visualizing the influence of a scholar over time. We apply the approach to the Microsoft Academic Graph, a large (publicly available) citation network.
We report on the development of this method through a design study with the Pew Biomedical Scholars program. 
We validate the design abstractions through demonstrations and discussions with the Pew program officers. 
We also report on the insights from a validation study in which the Pew scholars themselves interact with the visualization. We extend these methods to offer a publicly available service to visualize scholars' influence at \url{http://scholar.eigenfactor.org}. We conclude with a discussion of insights gained from this study and future opportunities for work in this space.


\hypertarget{background}{\section{Background}\label{background}} 

\subsection{Assessing Scholarly Influence}\label{assessing-scholarly-influence}

Communicating scholarship at the individual level for assessment has taken qualitative and quantitative forms. 
More qualitative methods include research narratives authored by a scholar herself, text articles written about a scholar, interviews, or the career retrospectives that occur at conferences and other scholarly events as a way of acknowledging the importance of scholars' contributions. These forms convey a scholar's career in detail in an accessible narrative form. 
However, these types of reviews take considerable time to prepare and do not easily scale.

Quantitative methods of capturing scholarly impact, often for evaluation purposes,
have been used for many years as well. These include measures such as counts of publications and
citations. The \emph{h}-index was introduced in 2005---a researcher's \emph{h}
is the maximum number \emph{h} so that \emph{h} papers have each been cited at least \emph{h} times
\cite{hirsch_index_2005}. Although this measure has received increased attention in recent years as a means of assessment, it still suffers from many problems of its predecessors, such as bias along academic field, academic age, and gender
\cite{kelly_h_2006, leydesdorff_caveats_2008}.

Another problem with methods that use straight citation counts is they do not take into account the quality of citations. Several methods
have been proposed that use the structure of citation networks to
algorithmically weight links according to their overall influence (a
method analogous to Google's PageRank for websites
\cite{page_pagerank_1999}); these include Eigenfactor \cite{West2013JASIST}, Y-factor,
CiteRank, and P-Rank \cite{yan_scholarly_2012}. Our approach employs the article-level
Eigenfactor metric, which ranks individual papers according to their
position in the network \cite{west_eigenfactor_2010}. 
However, while these methods are considered to be an improvement over
simple citation counts, in isolation they can still fall victim to similar issues and
biases.

We suggest that using visualization to convey scholarly impact can capture a scholar's influence in a way that provides both qualitative and quantitative information.  
Visualizations are often used as a means to engage novice and more expert users alike. Visualizations can make patterns and relationships in a data set clearer~\cite{larkin_why_1987},  and act as storytelling devices~\cite{segel2010}.
A well-designed visualization can also support analysis to varying degrees of detail, from providing a gestalt view of the overall pattern of a scholar's career while still allowing for more careful examination of the subtler differences in the type or magnitude of influence the scholar has had.

\subsection{Citation network visualization}\label{citation-network-visualization}

There is a large body of work on the topic of mapping and visualizing networks of scholarly publications. Many of the existing techniques define their links using similarity measures---bibliographic coupling, co-authorship, and co-citations. Relatively less work has been done visualizing direct paper-to-paper citation networks. (See \cite{cobo_science_2011} for a review.) Since we want to look at the \emph{influence} of scholars, we are more interested in these direct citation networks than measures of similarity.

There are several tools that do support visualizing direct citation networks, including Action Science Explorer \cite{dunne_rapid_2012}, the Network Workbench Tool \cite{borner_rete-netzwerk-red:_2010}, CitNetExplorer \cite{van_eck_citnetexplorer:_2014}, Citeology \cite{matejka2012citeology}, and PivotPaths \cite{dork_pivotpaths:_2012}. While some of these tools offer the ability to view a particular paper, including author selection, they are designed to support analysis of a network at a particular point in time. Our approach, in contrast, views a dynamic network over time to tell a story of changing and developing influence over the course of a career.

\subsection{Visualization of dynamic networks}\label{visualization-of-dynamic-networks}
Dynamic network depiction is a particularly challenging subset of network visualization due to the need to show changing structure while preserving the mental model of the user. 
Animation naturally affords interpretation of change over time~\cite{tversky}. 
However, to ease the cognitive cost of maintaining the mental model requires limiting change to node positions over time steps and/or smoothly interpolating node positions between frames~\cite{moody2005,purchase2007}. 
Our technique avoids the difficulty caused by changing node positions by using a radial layout with a fixed anchor point\cite{moody2005} from which new nodes (representing chronologically published papers by a scholar) spiral outward, encoding time redundantly with the animation.

Radial layouts have been used as a way to retain context by snapping nodes of interest to a central point to facilitate analysis centered on different nodes~\cite{yee2001}. 
Applications that map time to the distance from the center point are less common, though several static layouts are exceptions. 
TimeRadarTrees~\cite{burch2008} encodes changes across a sequence of graphs as circle sectors. Each circle sector extending outward from a center point represents a subsequent time step, and each sector is divided into as many sections as needed to depict the nodes and their incoming edges. 
TimeSpiderTrees also produce static visualizations of dynamic graphs using radial layouts, but by using orientation rather than connectedness to express relationships between nodes. The result is a sparser visualization in which half links between nodes represent changes~\cite{burch2010}. 
Farrugia et al. use a radial layout in which concentric circles represent time periods in dynamic ego-networks~\cite{farrugia2011}.
Radial layouts have also been used to depict an adjacency matrix at multiple times steps as rings of a circle~\cite{vehlow2013}.
We similarly use a metaphor of time as distance from the center of a circle to depict network data. However we use a spiral shape based on their ability to act as a metaphor for temporal change~\cite{aigner2011}.

\section{Methods}\label{methods}

\subsection{Context}\label{context}
We began exploring methods for visualizing scholarly influence
after being contacted in early 2015 by the Pew Scholars Program in the
Biomedical Sciences. This program provides four years of early-career
funding to approximately 30 researchers in health-related fields each year. They have funded multiple highly influential researchers in the biomedical sciences including several Nobel Prize winners. The program was celebrating its 30-year anniversary and wanted to reflect on its history using more than standard metrics alone (e.g., citation counts, h-indexes, impact factors). We met with the program directors to discuss richer ways of exploring their impact and influence on biomedicine.

The Pew Charitable Trust is one of many foundations and funding agencies trying to figure out how to measure their impact on scholarship. We viewed this evaluation as a case study in how to better visualize scholarly influence for individual scholars in general. The Pew scholars have been publishing for several decades, their publication data is readily available in open repositories like PubMed Central, and they tend to be influential scholars from a diverse set of disciplines. This prompted us to consider ways to convey not only how much influence the scholars have had, 
but also the qualitatively different kinds of influence that a scholar could have.

Based on the Pew program's goal of reflecting on their history and our own perception of 
a broader opportunity to use visualization to convey scholarly impact, we approached the design study as a case study in using data visualization as a storytelling device. 
Throughout the design study, we referred to the data on a scholar as a story comprising multiple events.
This emphasis on storytelling helped encourage us to explore ways of presenting the data that 
could make it accessible to users who are not necessarily accustomed to using interactive visualizations for analysis, in the same way that narrative visualizations are used to make data more accessible to audiences in the media and other outlets. 

\subsection{Design Study Methodology}
We developed the visualization by using an iterative design process over the course of about five months. We met remotely with the Pew program
officers eight times throughout this period.
Initial meetings consisted of discussing how to frame the
goals of conveying scholarly impact and to brainstorm specific measures and visual presentation styles (e.g., animation, static snapshots).
Subsequent meetings were used to demonstrate and receive feedback on the design iterations.
This process culminated in
demonstrations and testing with the Pew scholars at the reunion conference; this is discussed below in
\protect\hyperlink{results}{Results}.

We followed Munzner's nested model for visualization design and validation
\cite{munzner_nested_2009}. This model characterizes visualization
design and evaluation at four nested levels---problem characterization,
data/operation abstraction, encoding/interaction, and algorithm---and
identifies threats to validity at each level. 
In the next section, we
address the last three levels, describing our design process and addressing threats to validity through justification or evaluation.

\section{Design}\label{design}

\begin{figure*}[tbp]
\centering
\includegraphics[width=17cm] {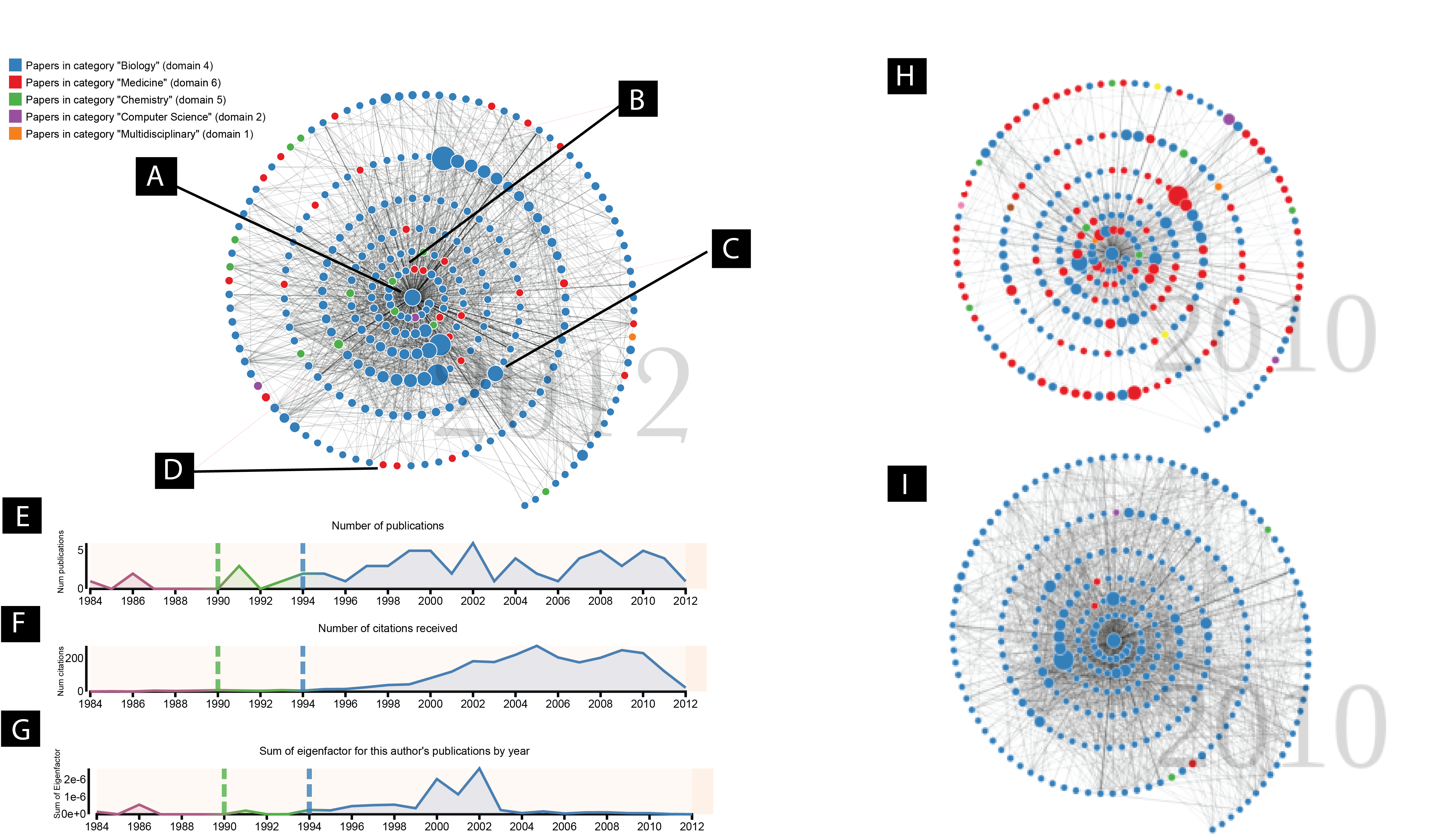}
\caption{\emph{Top Left:} (A) The center node represents all publications of a particular scholar. (B) Nodes that appear around the center represent publications that cited work by this scholar. (C) The size of the nodes show a citation-based indicator (Eigenfactor) of how influential that paper has been. (D) Colors show different fields to which the papers apply. \emph{Bottom Left:} Integrated timeline charts below the network visualization.
(E) Number of publications by the central scholar by year. (F) Number of citations received by the central scholar by year. (G) Sum
of the Eigenfactor for all of the publications published by the central
author in each year. Colors show the periods before, during, and after funding from the Pew program. \emph{Right side:} Comparing the densities of two different graphs. (H) is a sparse graph that shows a diffuse influence across fields (i.e., interdisciplinary influence). (I) is a dense graph that shows a close-knit citation community within one domain.}
\label{fig:annotatedVis}
\end{figure*}

\hypertarget{data-abstraction}{\subsection{Data abstraction}\label{data-abstraction}}

\textbf{Data set}: Our database of scholarly publications comes from a public release of Microsoft Academic Search. The data set for our study contains about 49 million papers and 260 million citations. Papers have associated metadata such as title, year, list of authors, journal or conference, abstract, etc. There is also an assigned domain for each paper (e.g. ``Biology,''
``Chemistry,'' ``Computer Science'')---this domain has been assigned by Microsoft at the time of collection. Since the initial design study, we have switched to the February, 2016 release of the Microsoft data (available at \url{https://www.microsoft.com/en-us/research/project/microsoft-academic-graph/}); this updated data set has about 127 million papers and 528 million citations \cite{sinha_overview_2015}.

\textbf{Graph representation}: 
We represent scholarly publications as nodes in a network, and citations as directed links between them. Additional features relevant to assessing influence are stored as node attributes. These include year of
publication, title, authors, and domain. 
The Eigenfactor score---a metric of influence for each paper that takes into account the number and quality of its citations, calculated across the entire data set---is
also stored as an attribute of each node (see~
\protect\hyperlink{background}{Background} above).

We transform the data into a directed \emph{egocentric network,} a
subset of the total graph that considers one central node (the ego) and
all of its neighbors (the alters), as well as all of the edges from
alters to ego and between alters \cite{butts_social_2008}.
The center node represents the set of all papers authored by a particular scholar.
This approach requires author identification as a subtask: determining which papers in a large scholarly database are authored by a given individual (see
\protect\hyperlink{implementation-algorithm}{Implementation/Algorithm}
section below).

Taking this approach, the center (ego) node represents the total body of
work authored by the scholar being visualized (Figure~\ref{fig:annotatedVis} A). All of the scholar's
papers and their associated features are stored as attributes on the
center node. All of the alter nodes represent individual papers that
have cited any of the papers contained in the ego node. The alters all
have at least one link to the center node---multiple if the paper cited
more than one paper authored by the scholar of interest---as well as
links to other papers that appear in the egocentric network.  The Pew scholars we visualize have some variation in the total number of nodes in their network, typically between around 200 and 5000. As described below in \protect\hyperlink{implementation-algorithm}{Implementation/Algorithm}, we limit the number of nodes displayed in the graph portion of the visualization (n=275 for all figures in this paper) in order to keep the level of visual complexity under control.

\textbf{Key Indicators over Time}:
Additional data transformations calculate key indicators of the scholar's career over time. Each of these indicators are calculated for each year: total number of publications authored by the scholar, total number of citations received by any of the scholar's papers, and sum of the Eigenfactor influence scores for all of the papers authored by the scholar in each year. Since we use the Eigenfactor score as a measure of the citation-based influence of an individual paper, the Eigenfactor score sum can
be thought of as a measure of the total
(citation-based) influence the scholar's output has had that year. Each of these
indicators contextualizes the career-level data from a
different angle. These indicators are visualized over time in linked timeline charts that appear below the graph display (See Figure~\ref{fig:annotatedVis}--E, F, and G).

\textbf{Validation}:
The most important data abstraction decision to validate is our conceptualization of influence. Through discussions with the Pew officers, and informed by a broader awareness of how influence can be conceptualized, we identified features in our data that reflected different facets of influence.  Measures of citation counts and importance of publications in the overall network (i.e. Eigenfactor) show clear but rough indications of the influence a scholar has had. Features such as the domain of the citing work and the number of connections citing papers have had to other citing papers say something about the type of influence the scholar has had---whether it tends to be concentrated in a small community or diffuse to different areas.

Additional downstream validation of the data abstractions came through testing the design with Pew scholars (see \protect\hyperlink{results}{Results}).

\hypertarget{visual-encoding}{\subsection{Visual encoding and interaction
design}\label{visual-encoding}}

The graph is represented visually in the common paradigm of the
node-link diagram, with circular nodes representing vertices connected
by straight lines representing the edges between them (Figure~\ref{fig:annotatedVis}--A-D). The ego node, representing all of the central scholar's papers, is
placed at the center of the display (A). The alter nodes, representing papers citing the ego's work, surround
the ego node (B). 
All nodes and links are hidden initially, then are animated in chronological order by year, extending in a spiral layout from the center node, beginning with the year of the earliest publication by the central scholar. A year counter behind the graph displays the publication year of the nodes and links currently appearing.
The direction of the links is encoded by
the animation: links are sent out from each citing node to the cited
nodes after the citing node appears. The rate at which nodes appear is based on
the number of nodes in each year, so that nodes appear more quickly in
years with more citing; this is meant to lend more excitement to the
more active years. If an alter node has more than one link to the center node (i.e. the paper has citations to multiple papers authored by the ego scholar), multiple lines are drawn on top of each other, so that edge thickness is mapped to number of citations. While the final node-link structure is often complex and interpreting the meaning of individual links is difficult, the intent is to convey a high-level view of the connections that form around a scholar in her citation community, and to allow relative comparisons of density. More focused analysis is supported by details on demand for a citing paper via mouseover of nodes. The user can also click on a node to be taken to either the full text of the paper (if available) or the paper's Microsoft Academic page.

\textbf{Animation}: The use of animation to show the network build over time was an important design choice throughout the process. A goal of this visualization was to use the data to tell a story that would be compelling to a wide audience. While it can 
have drawbacks, animation as a medium naturally draws attention and can encourage perceptions of narrative~\cite{palmer}. 
By using animation to encode time series data as observable changes, metaphoric change may also be communicated~\cite{tversky}.

\textbf{Spatial encoding}: We experimented with multiple spatial encodings of the nodes. Initial designs used a force directed layout commonly used in node-link diagrams, to place the alter nodes around a fixed ego. This placement, however, tended to produce an
overwhelming visual representation that was difficult to interpret (the
``hairball'' effect also commonly associated with node-link diagrams). It also did not make effective use of spatial
placement as means of encoding something useful about the data.

We chose to place the nodes in a spiral pattern for several reasons. By
ordering the nodes by year and placing them outward from the center, it
allowed us to encode temporal information in the network---increased
radial distance represents a more recent publication date, one that is
later in the scholar's career. The original force-directed layout
encoded publication date only temporally, with earlier dates being
revealed earlier in the animation. The spiral layout adds the spatial
encoding to reinforce this dimension, making the narrative easier to
follow. Another advantage of the spiral placement was the ability to include more nodes in a limited space without too much overlap and confusion. The tradeoff of this placement is that it precludes optimizations intended to minimize edge crossing.

\textbf{Other encodings}: We chose to encode two additional features on the network's alter nodes: Eigenfactor and domain. The Eigenfactor of each paper is represented by the relative size of the node (Figure\ref{fig:annotatedVis} C). This allows the viewer to easily identify some of the most important papers that have cited the center scholar's work (see \protect\hyperlink{background}{Background} section above for more about Eigenfactor). The domain of each paper is represented by the color of the node (Figure\ref{fig:annotatedVis} D), with a legend generated for each scholar on the top left of the display identifying which colors map to which domains. The most common domain among the papers in the ego node is set as blue, and other domains that appear in the network are assigned to a categorical color scheme in order of frequency with which they appear in the graph. We chose a relative rather than absolute color scheme because there exist too many fields to assign each a color. In addition to showing individual papers from different fields, the extent of color variation in the total graph allows the viewer to see at a glance the extent to which the influence of a scholar's work tends to cross intellectual boundaries.

\textbf{Timeline visualizations}: 
Three timeline charts appear below the graph. The x-axis shows the years from the earliest paper authored by the center author to the last year in our data set for which we have data. Figure~\ref{fig:annotatedVis}--E, F, and G show the timeline charts; see the \protect\hyperlink{data-abstraction}{Data abstraction} section above for a description of the data abstractions shown. As the time progresses, the current year being visualized is highlighted in the timeline charts in orange. The years that have already been visualized are highlighted in faint orange. The viewer may click on a year in the x-axis to move the animation forward or backward to the state of that year. One additional dimension was encoded for the interest of the Pew program officers---colors and vertical lines show the periods before, during, and after the funding that the Pew program provides to the scholars. This is one example of how overlaying additional data can help to add context to the overall story, and is discussed more in the \protect\hyperlink{future-work}{Future Work} section below.

\textbf{Comparing visualizations}: The scales used throughout the visualization---the mapping of Eigenfactor to node size, the color of the domain, and the y-axes on the timeline charts---are calculated relative to each individual scholar. This makes comparisons between different scholars on these dimensions difficult. This was a deliberate design choice. As discussed in \protect\hyperlink{background}{Background} above, quantitative metrics exist and are already widely used to compare scholars based on measures of output and citation counts. Our initial intent in working with the Pew program was to discourage comparison and ranking in favor of a more qualitative view of an individual scholar's influence.
However, as we generated visualizations for
different scholars, we did notice certain patterns that said something
about the different types of influences.
The right side of Figure~\ref{fig:annotatedVis} shows two different graphs, one dense
and monochrome (H), the other sparse and colorful (I). One is not necessarily
more influential than the other; rather, they exemplify two different
ends of a spectrum of influence, which can be seen in the citation
pattern around the scholars' work. The dense, monochrome graph
shows a scholar who tends to have influence in a specific area, a
close-knit group of researchers that have many connections to each
other. The sparser, more colorful graph shows a scholar who has had
diffuse influence in different disciplines. The papers that cite this type of
scholar tend to cite other papers that appear in the graph less often,
resulting in fewer links between alters and a sparser network. Supporting these types of comparisons will be important as we continue to develop these methods (see \protect\hyperlink{future-work}{Future Work})

\hypertarget{implementation-algorithm}{\subsection{Implementation/Algorithm}\label{implementation-algorithm}}

Implementing the overall design is carried out in several stages: identifying the author in the database, collecting and caching the data, and drawing the visualization.

\textbf{Author identification}: 
Inaccurate \emph{author disambiguation} is a threat to the validity of the depiction of scholarly impact. 
A unique identifier in the data set corresponds to an author identified by the collection algorithms; however, a single scholar may actually correspond to several IDs, and scholars with common names may be mistaken for different
people due to inaccuracies in the algorithms Microsoft uses to collect the data.
To mitigate the potentially misleading view of influence that can occur from disambiguation errors upon inputting only an author name, user input is required. The latest version of the system---hosted on \url{http://scholar.eigenfactor.org}, allows users to curate their own collections of papers, selecting and removing papers from the collections as they see fit before generating the data and visualization.

\textbf{Obtaining and Representing Data}:
The next stage of implementation is putting the data (stored in a MySQL
database) into a network structure using the Python packages pandas
\cite{mckinney_data_2010} and NetworkX \cite{hagberg_exploring_2008}.
Starting with a graph with the ego node representing a
scholar, the total set of papers associated with this scholar (as curated by a user) are stored as an attribute of the ego node. For each of these papers, the citing papers are collected and added to the graph as an alter node. Finally, for each citation by an alter paper, an edge is created between alter and ego if the cited paper is in the ego node, or between alter and alter if the cited paper appears in the graph.

\textbf{Visualization Rendering}:
The final stage of implementation is the visualization, implemented
using the open-source JavaScript library D3 \cite{bostock_d3:_2011}.
For the network representation, in order to reduce the visual
complexity, the number of total nodes is capped at 275; if there are
more, the alter nodes are chosen based on Eigenfactor and whether they have associated domain information. The alter nodes are then sorted by year, placed in a spiral formation around the center, and hidden. The speed at which nodes appear is calculated based on the number of nodes in the current year being animated, using a threshold-based scale that sets the total time per year. This scale is set to achieve a balance between smooth narrative and having nodes appear faster in years with more activity. Years with very few nodes take .8 seconds, while years with 30 or more nodes take 4 seconds to animate (with multiple threshold settings in between); empty years take .3 seconds.

The number of nodes to visualize (n=275) and the spacing of the nodes is hard-coded, and was arrived at after some trial and error. The goal was to show as many nodes as possible in the space typically afforded by a web visualization, while avoiding excessive overplotting and occlusion. We arrived at this design after going through several iterations in collaboration with the Pew officers.

\hypertarget{results}{\section{Results}\label{results}}

\subsection{Evaluation with Scholars}\label{evaluation-with-scholars}

The Pew program held a three day meeting in November 2015 for their 30th reunion
with approximately 400 scholars attending, ranging from the first class of 1985
to the class of 2011 (a scholar's Pew class is the year that he or she
was accepted to the program and began to receive the four years of
funding). Throughout the three days, the scholars attended research
talks and social events. We set up a table with a display so that
scholars could view and interact with the visualization during their
down time. When a scholar approached the table, we demonstrated the visualization
with her data and allowed her to watch and interact. We then asked
open-ended questions to prompt a dialog. 

During the reunion, we
demonstrated the visualization with 26 scholars. We also allowed the
scholars to access the visualization on their own online, and encouraged them
to contact us with any feedback. Since the demonstrations, we have received approximately 20 emails and engaged in 15 informal conversations providing additional feedback. In this section, we discuss high-level observations that emerged from these demonstrations and conversations. In the next
section, we present several interesting individual stories that
came out of the experience.

While interest in viewing and interacting with the visualization was
high, many of the scholars approached with skepticism. 
Many scholars are wary of the limitations of evaluations based solely on publications, 
and a common frustration expressed among the Pew scholar was the use of measures
such as citation counts and h-index. However,
we observed that for most scholars reactions shifted to a generally positive tone after trying out the visualization.
While the concerns were not completely assuaged, we believe that the scholars tended to
appreciate how the visualization represented different dimensions of
influence to present a richer picture than these common metrics.
Several scholars noted this aloud.

Several scholars struggled with the fact that nodes represented citing papers, rather than the scholar's own papers.
We suspect that this difficulty adjusting to nodes representing citing papers may be partly due to the emphasis on the individual scholar's papers in many current scholarly databases that offer ego-views, such as individual scholar's DBLP or Google Scholar profiles. An interesting avenue for future work is to integrate a depiction of the scholar's own papers as part of the visualization.

Another common issue with the data abstractions that came up during
these validations was that of the difference between review articles and
original research. Review articles tend to be highly cited papers,
especially in the biomedical field, and thus may be overrepresented in
the graph display. When the scholars interacted with the visualization
and began identifying some of the larger nodes, they found that many of
them were in fact review articles. While many scholars agreed that it
was noteworthy to be cited by a prominent review article, some thought
that review articles represented something different from original
research, and thought that the distinction should be made clear. These comments made us aware that the influence of review articles can be a contentious topic among 
some scholars, who believe that they should be omitted entirely from influence measures.
Future work can focus on making this distinction clearer, 
and devising ways of identifying which papers in the network scientists tend to
consider more important and influential.

Most of the scholars were interested in comparing their data to others',
asking tentative questions along the lines of, ``Is my spiral good?'' As discussed
above in the \protect\hyperlink{visual-encoding}{Visual Encoding} section, we tried to discourage these sorts of
comparisons. While it is possible to see absolute differences between
scholars--for example, by examining the scales in the y-axes of the
timeline charts to see who had more publications or citations or comparing the density of the link structure across graphs (Figure~\ref{fig:annotatedVis} H, I)---the visualization is not designed to make these differences prominent. 
Our intent was to use our
data to highlight the different types of influence these researchers
have had, and it was usually possible, with some effort, to steer the
focus in that direction.

\subsection{Stories from the Scholars}\label{stories}

One of the most interesting results to come out of the demonstrations
was that viewing their data frequently prompted the scholars to reflect on
their careers and to tell stories about how what they saw on the screen
matched up with how they saw their own histories. There were many
comments about how certain peaks or dips in the timeline charts, or
changes in activity or color on the graph, corresponded to career
shifts, restructuring of laboratories, or even significant personal
events.
The visualization thus served as a catalyst for communication around a particular scholar's trajectory, at some points fostering 
discovery by the scholar of influences and dynamics in their career of which they had not been aware. In this section, we present several specific stories that
emerged.

\begin{figure*}[bt]
\centering
\includegraphics{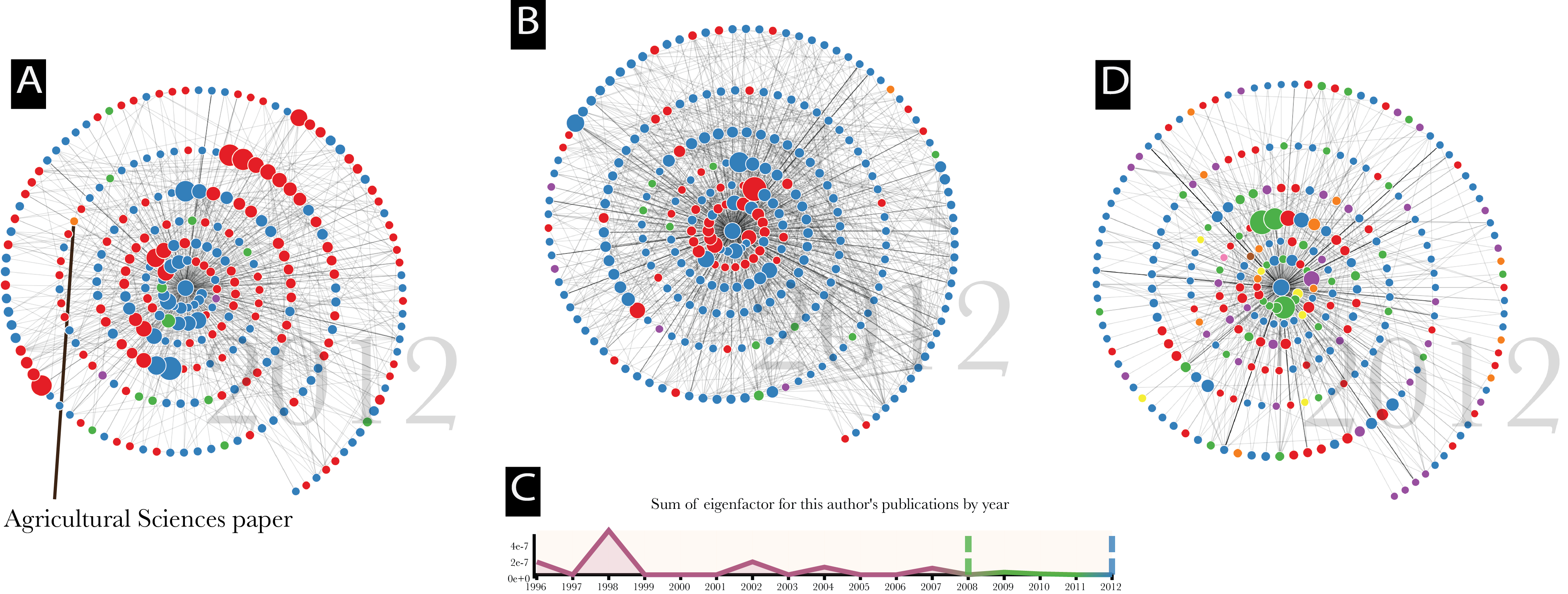}
\caption{Four stories that emerged from demonstrations with the scholars. A) shows a scholar who had influence in a field she hadn't expected. B) shows a career shift reflected in changing color bands in the graph. C) shows an early-career peak in influence that prompted a scholar to reflect on the freedoms afforded by different research positions. D) shows a scholar with influences in very diverse areas.}
\label{fig:pewStories}
\end{figure*}

One scholar, when shown her citation network, noticed that she had been
cited by a prominent paper in the Agricultural Sciences literature (Figure~\ref{fig:pewStories}A). At
first she identified this as an error in the data. Her area of study is
the cellular mechanisms underlying heart attack, and she didn't see
herself as having any connection to the study of agriculture. However,
on further reflection, she made the connection that a particular protein to which
she had devoted a period of her career was also involved in meat
tenderization. In this case, the self-reflection enabled by visualizing
this scholar's citation network enabled her to identify an influence she
had had on a completely different field, one which she hadn't considered
before.

Another scholar's graph showed a dramatic shift in color from the inner to the outer nodes of the spiral (Figure~\ref{fig:pewStories}B). Talking to this scholar, he agreed that this reflected a major shift in his career, when the focus of his research changed from a topic in chemistry to one in biology. The papers that tended to cite his work changed as well, and the color patterns in his graph conveyed this shift in influence in a way that was easily recognizable to him.  We also saw these shifts in color when scholars changed model organisms (e.g., \textit{Arabidopsis} to \textit{Drosophila}).

These methods do not tend to work as well for young scholars, as a longer career provides more input data for telling the story of their developing influence. Nevertheless, one young researcher noticed a peak in her Eigenfactor timeline chart that corresponded to some of her work in graduate school (Figure~\ref{fig:pewStories}C). This led her to reflect on her time in graduate school and the boundary-pushing research that she conducated at that time. Her current research position, she said, allows much less of this type of freedom. In this case, the scholar was able to imbue the data visualized on her chart with her personal story of how she felt about her research's ability to have meaningful influence.

One scholar, before viewing his visualization, jokingly commented on how unfocused he was---he tended to publish on a wide
range of topics and expected his citation graph to reflect this. As
promised, his graph turned out to be the most colorful we had seen,
reflecting a career whose influence had reached researchers in
chemistry, medicine, biology, material science, engineering, physics,
computer science, and environmental science (Figure~\ref{fig:pewStories}D). The alter nodes in his
graph do not have many connections to each other, which is another
indicator that his influence has reached a diverse set of communities.
This scholar enjoyed seeing his story visualized in this way, and wanted
to feature the graph on his personal website. 
He also used the visualization as a chance to reflect on his future plans, mentioning that he
expects the graph to get ``even worse''---i.e.~more colorful and
reflective of more diffuse influence---in the future as 
citations stemming from his recent work increase.

\hypertarget{future-work}{\section{Discussion and future work}\label{future-work}}

While our focus in this work was developing narrative visualizations for the Pew scholars, we have already begun to use it to generate visualizations of scholars outside of the program. We have also applied the methods to entire fields of study rather than individual scholars~\cite{Portenoy2016www}. As we continue in this direction, future work will address the generalizability of all of our design choices---whether, for example, it might be better to use the data to choose the proper number or spacing of nodes, rather than having this predetermined. Identifying which nodes to display is also a question we will revisit, as we reconsider which papers in a scholar's network are most salient to show influence. To do this, we will ask the Pew scholars themselves to note their most influential papers. This was an idea from one of the Pew scholars.  

Our goals in working with the Pew program centered around creating visualizations to help the program reflect on its set of highly influential scholars. 
This shifted our focus away from direct comparisons of different scholars. As we broaden our scope and generalize out to include scholars outside of the program, however, one of the most important directions for future work will involve turning more toward comparison---addressing the question of how to place one author's story in a larger context. Displaying two visualizations side by side would be one option, with an author's display appearing next to an appropriate control. Thought needs to go into selecting these controls---for example, an aggregated representation of other authors with similar careers, or, in the context of evaluating impact for funding agencies, an aggregated control based on scholars who applied for funding but were not awarded or did not accept funds.

Another direction for future work relates to the narrative nature of the visualization: how to incorporate different types of data into the story told by the animation. We have shown one example of overlaying additional data to deepen the context: the coloring and labeling of the timeline charts by Pew funding period. Adding this dimension helped the Pew officers and scholars to reflect on the stories and consider the effect that entering into and receiving funds from this program may have had. Other additional encodings could support program and individual evaluation in a number of other settings.

Other forms of data could also be integrated to further emphasize the visualization as a storytelling device. 
Automated annotation of salient shifts in the magnitude or domain of influence could help guide a novice user's interpretation.
Multimedia storytelling through the integration of audio is another interesting avenue for future work. The Pew program, for example, has conducted interviews with most of its scholars and has both audio and transcripts available. Excerpts from these interviews, played at the proper time during the animation, could provide additional dimensions to the overall story of the scholar's career.

Expanding out from the case study with the Pew scholars, the website \url{http://scholar.eigenfactor.org} will serve as a launching pad to offer as a free service this and other tools to analyze and visualize scholarly influence using citation graphs. User data and feedback will be helpful in expanding and developing these tools.

\hypertarget{conclusion}{\section{Conclusion}\label{conclusion}}
We presented a design study in the domain of visualizing scholarly influence to tell a scholar's story, collaborating with the Pew Biomedical Scholars program and using their scholars as an initial case study. We described our design process of choosing data abstractions and visual encoding techniques in collaboration with Pew program officers, and detailed the implementation. We demonstrated the visualization with the scholars, and identified general trends and specific stories that showed how the visualization helped the scholars to reflect on their own influence. Finally, to generalize the methods to more scholars, we implemented a system which allows users to curate collections of papers and generate visualizations themselves.

\hypertarget{acknowledgements}{\section{Acknowledgements}\label{acknowledgements}}
We would like to thank the Pew Charitable Trust and the Chemical Heritage Foundation for funding, and for allowing us to interact with the program managers and the Pew Biomedical Scholars. We also want to thank Microsoft Research for providing the citation data through their Microsoft Academic Graph.

\bibliographystyle{acm}
\bibliography{authorvis}

\end{document}